\documentclass[a4paper,12pt]{article}  
\usepackage[dvips]{graphicx}
\usepackage{amssymb}

\newcommand{\half}{{\textstyle{1\over 2}}} 
 
\newcommand{\goes}{\rightarrow} 
 
\newcommand{\GeV}{\; \mathrm{GeV}} 
\newcommand{\TeV}{\; \mathrm{TeV}} 
\newcommand{\beq}{\begin{equation}} 
\newcommand{\eeq}{\end{equation}} 
\newcommand{\bea}{\begin{eqnarray}} 
\newcommand{\eea}{\end{eqnarray}}

\newcommand{\strahl}{e^+e^- \goes Z h^0 \goes \bar{\nu} \nu \, h^0}
\newcommand{\wwfus}{e^+e^- \goes \bar{\nu}_e \nu_e \,  W W 
                                  \goes \bar{\nu}_e \nu_e \,  h^0}

\newcommand{\higpro}{e^+e^- \goes \bar{\nu} \nu h^0}

\begin{document} 
\begin{titlepage} 
 
\begin{flushright} 
HEPHY-PUB 754/02 \\  
hep-ph/0204280  
\end{flushright} 
\begin{center} 
\vspace*{1.5cm} 
 
{\Large{\textbf {Radiative corrections to single 
    Higgs boson production in \boldmath{$e^+e^-$} annihilation}}}\\ 
\vspace*{10mm} 
 
{\large H.~Eberl, W.~Majerotto, and V.~C.~Spanos} \\ 

\vspace{.7cm} 

{\it Institut f\"ur Hochenergiephysik der \"Osterreichischen Akademie
der Wissenschaften,}\\
{\it A--1050 Vienna, Austria}

\end{center} 
\vspace{3.cm} 
\begin{abstract} 
For energies relevant to future
linear colliders, $\sqrt{s} \gtrsim 500 \GeV$, the $WW$ fusion
channel dominates the Higgs boson production 
cross section $e^+ e^- \goes \bar{\nu} \nu h^0$. 
We have calculated the one-loop corrections to this
process due to fermion and sfermion loops in the context of the MSSM. 
As a special case, 
the contribution of the fermion loops in the SM
has also been studied. 
In general, the correction is negative and sizeable of the order of $10\%$,
the  bulk of it being due to fermion loops. 
\end{abstract} 
\end{titlepage} 

\baselineskip=18pt

As no Higgs boson could be detected so far, the search for the Higgs
boson as the primary goal of high energy physics will continue.
The four LEP experiments delivered a lower bound for
the Standard Model (SM) Higgs mass, $m_h \gtrsim 114 \GeV$ \cite{higgs}. 
In $e^+e^-$ collisions,
for energies $\gtrsim 200 \GeV$, the production of a
single Higgs boson plus missing energy starts to be dominated by
$WW$ fusion \cite{fusion,alta,kilian}, 
that is $\wwfus$, whereas the Higgsstrahlung process \cite{ellis}
$\strahl$ becomes less important.
$h^0$ denotes the SM Higgs boson or the light Higgs boson of the 
Minimal Supersymmetric Standard Model (MSSM), 
while $H^0$ is the heavy $CP$-even Higgs boson in the MSSM.
The rates for the $ZZ$ fusion are generally one order of
magnitude smaller than those of the $WW$ channel.

At LHC, in $p\,p$ collisions, the gluon-gluon fusion mechanism provides the
dominant contribution to Higgs production. The next important
Higgs production channel is vector-boson fusion $VV \goes h^0 / H^0 $.
In particular, it provides an additional event signature due
to the two energetic forward jets. Recently,
it has been argued that the channels 
$WW \to h^0/H^0 \goes \tau\bar{\tau}$ 
and $WW$, can serve as suitable
search channels at LHC 
even for a Higgs boson mass of $m_h \sim 120\GeV$  \cite{zepe}.

At Tevatron, 
with $p \, \bar{p}$ collisions at $2 \TeV$, 
the $WW$ fusion process plays a
less important r\^ole. The $g\, g$ fusion is the dominant
process for Higgs production there, but
the Higgsstrahlung $q \, \bar{q} \goes W \goes W h^0 $ is
larger than the $WW$ fusion for $m_h \lesssim 180 \GeV$ \cite{tevatron}.

It is worth mentioning that $WW$ (and $ZZ$) fusion is also the most
important Higgs boson production mechanism in $e\, p$ collisions.
Actually, it was shown \cite{aachen} that the $e\, p$-option at LHC 
would offer the best opportunity to search for a Higgs
boson in the mass range $m_h < 140 \GeV$.

In this paper, we have calculated the leading one-loop corrections
to the $WWh^0$ vertex in the MSSM by 
taking into account  fermion/sfermion
loops. They are supposed to be the dominant corrections due to the 
Yukawa couplings. We have applied them to the Higgs boson production in 
$e^+e^-$ annihilation in the  energy range
$\sqrt{s}=0.5 - 3 \TeV$, i.~e. $\wwfus$.
We included the Higgsstrahlung process $\strahl$ and the
interference between these two mechanisms.
Because the Higgsstrahlung process is much smaller in this range,
we have neglected its radiative corrections.
We  have also discussed the SM case where we have studied the
dependence on the Higgs boson mass. As to the one-loop corrections to the
$WWh^0$ vertex in the SM, we quite generally refer to the review article
\cite{kniehl-review}. For this coupling also QCD corrections were
included, the ${\cal O}(\alpha_s\,G_F\, m_t^2)$ in \cite{kniehl1}
and ${\cal O}(\alpha^2_s\,G_F\, m_t^2)$ in \cite{kniehl2}.

\begin{figure}[t]  
\centering\includegraphics[scale=1.]{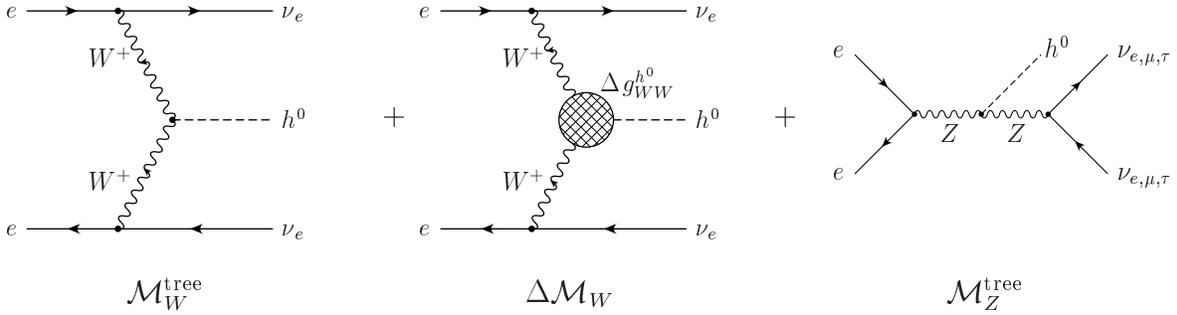} 
\caption[]{ The Feynman graphs for the process $\higpro$.
Note that for $|{\cal M}_Z^{\rm tree}|^2$ one has to sum over all
three neutrino flavors.} 
\label{fig1}
\end{figure}

As for energies $\sqrt{s} > 500 \GeV$ the dominant channel $\higpro$
is by far the $WW$ fusion, in the context of our calculation
only corrections to the $WWh^0$ vertex must be evaluated,
see Fig.~\ref{fig1}.
At the one-loop level the Lagrangian for the  
$WWh^0$ coupling can be written as
\beq
{\mathcal L}=\left( g^{h^0}_{W W} \,g^{\mu \nu}
 +   \left(\Delta g_{WW}^{h^0}\right)^{\raisebox{-1.3mm}{$\scriptstyle \mu\nu$}}
 \right) \, 
       h^0 \, W^+_\mu \, W^-_{\nu}\, ,
\label{lagr}
\eeq
where the tree-level coupling in the MSSM can be read as
$ g^{h^0}_{W W} = g \, m_W \, \sin(\beta - \alpha )$.
The one-loop part of the $WWh^0$ coupling in Eq.~(\ref{lagr}) can be
expressed in terms of  all possible form factors\footnote{As the
expressions for $F^{00}$, $F^{11}$,...,$F^\epsilon$ are
quite lengthy they will be presented elsewhere \cite{EMS}.} as
\beq
   \left(\Delta g_{WW}^{h^0}\right)^{\raisebox{-1.3mm}{$\scriptstyle \mu\nu$}} =
   F^{00} g^{\mu\nu} + F^{11} k_1^\mu k_1^\nu 
 + F^{22} k_2^\mu k_2^\nu 
 + F^{12} k_1^\mu k_2^\nu 
 + F^{21} k_2^\mu k_1^\nu  
 + i\,F^\epsilon \epsilon^{\mu\nu\rho\delta} 
                         k_{1\rho} k_{2\delta} \, ,
\label{ff}
\eeq
where $k_{1,2}$ denote the four-momenta 
of the off-shell $W$-bosons. 
At tree-level only the 
structure with $g_{\mu \nu}$ is present, and therefore
all form factors but $F^{00}$ have to be 
ultra violet (UV) finite without being renormalized. 
By adding appropriate counter terms also the form factor $F^{00}$
is rendered UV finite.
For the renormalization procedure the on-shell scheme has been
adopted.

\begin{figure}[th]  
\centering\includegraphics[scale=.82]{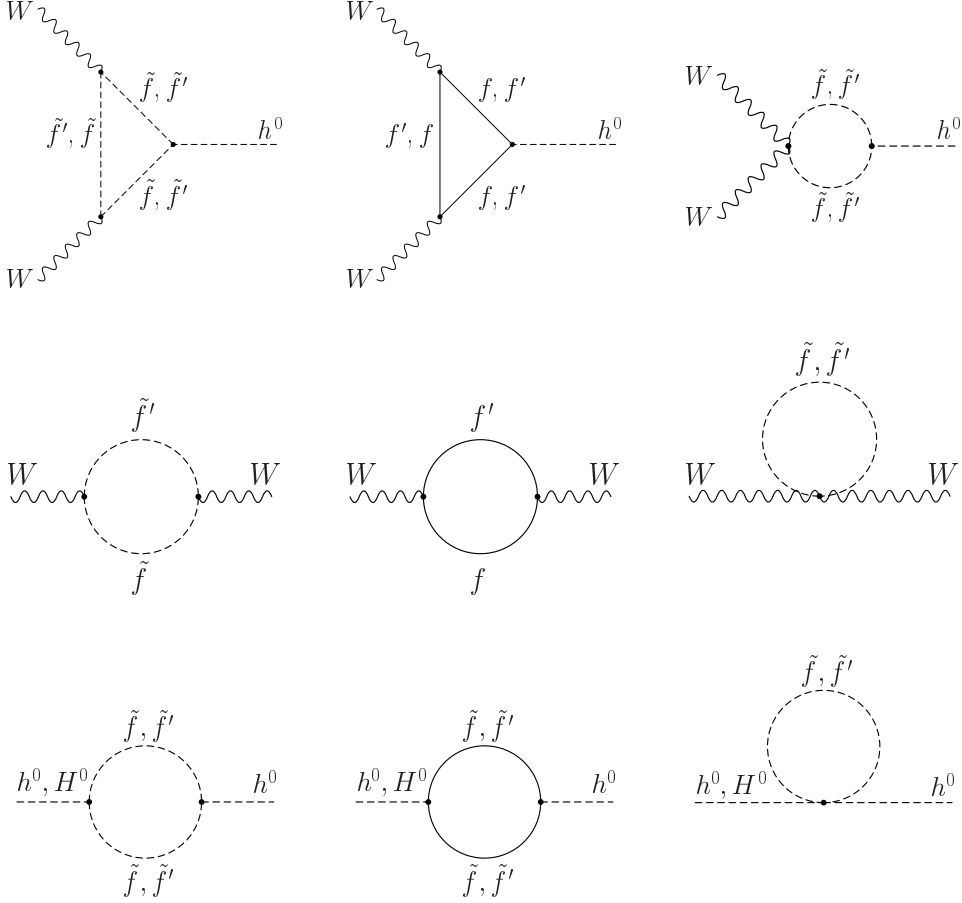}  
\caption[]{The Feynman graphs that contribute to the vertex 
corrections defined in Eq.~(\ref{ct}) and the
wave-function in Eq.~(\ref{wave}). 
 $f$ ($f'$) denotes the up (down) type fermion.} 
\label{fig2}
\end{figure}

Actually, for the calculation of the one-loop corrected 
$WWh^0$ vertex we compute the vertex and the wave-function
corrections stemming from the graphs of Fig.~\ref{fig2}, as
well as the coupling correction, 
\beq
 \left(\Delta g_{WW}^{h^0}\right)^{\raisebox{-1.3mm}{$\scriptstyle \mu\nu$}} =
 \left(\delta g_{WW}^{h^0\,(v)}\right)^{\raisebox{-1.3mm}{$\scriptstyle \mu\nu$}} +
\left(\delta g_{WW}^{h^0\,(w)} 
   +  \delta g_{WW}^{h^0\,(c)}\right) g^{\mu\nu} \, .
\label{ct}
\eeq
Comparing Eq.~(\ref{ct}) with Eq.~(\ref{ff}),
it is obvious
that the counter terms, i.~e. the wave-function and coupling corrections, 
contribute only to  $F^{00}$.
The wave-function correction is 
\begin{equation}
\delta g_{WW}^{h^0\,(w)} = g_{WW}^{h^0}\, \left( \half \, (\delta Z_H)_{h^0 h^0} 
 + \delta  Z_W\right) + \half \, g_{WW}^{H^0}\, (\delta Z_H)_{H^0 h^0}  \, ,
\label{wave}
\end{equation}
where $\delta Z_H$ and $\delta  Z_W$ 
are the symmetrized Higgs boson and the \mbox{$W$-boson} wave-function
correction, respectively, 
calculated from the corresponding graphs
of Fig.~\ref{fig2}.
In the case of the off-shell $W$-bosons, $\delta  Z_W$ has the form
\beq
\delta Z_W
   = \frac{\delta m_W^2 - \Re\Pi_{WW}^T (k_1^2)}{k_1^2 - m_W^2} 
+ \frac{\delta m_W^2 - \Re\Pi_{WW}^T (k_2^2)}{k_2^2 - m_W^2} 
+ 2\, \frac{\delta g}{g}\, .
\label{dzw}
\eeq
The coupling correction is 
\beq
\delta g_{WW}^{h^0\,(c)}  =  
\left( \frac{\delta g}{g} + \frac{\delta m_W}{m_W}\right) g_{WW}^{h^0} 
           + \frac{\sin2\beta}{2}\,
\frac{\delta \tan\beta}{\tan\beta} \, g_{WW}^{H^0}\, . 
\label{dgc}
\eeq
The expressions on the right-hand 
sides of Eqs.~(\ref{wave}--\ref{dgc}) 
can be found in Ref.~\cite{EMKY}.
Especially, we fixed the counter term $\delta\tan\beta$ by the
on-shell condition ${\textrm{Im}} \, \hat{\Pi}_{A Z}(m^2_A)=0$,
where  $\hat{\Pi}_{A Z}(m^2_A)$ is the renormalized self-energy
for the mixing of the pseudo-scalar Higgs boson $A^0$ and $Z$-boson.

Having calculated the form factors of  Eq.~(\ref{ff}),
one can proceed to the calculation of the one-loop corrected
cross section. The tree-level part of the amplitude was
already calculated in Ref.~\cite{kilian}. Here we will include
also the one-loop correction, that is the interferences between
the tree-level and the 
one-loop amplitudes of Fig.~\ref{fig1}\footnote{As it has already been
mentioned the corresponding corrections to the Higgsstrahlung
process can be safely ignored due to absolute dominance of the
$WW$ fusion for $\sqrt{s}> 500 \GeV$.}.
For these terms we get
\bea
  2\, \Re\left[\Delta {\cal M}_W 
     \left({\cal M}_W^{\rm tree}\right)^\dagger\right] & = &  
  4\, g^{h^0}_{W W} g^4\, 
 \left(2\,F^{00}\, p_1 \cdot p_4 \, p_2\cdot p_3 
 + F^{21}\, S\right) \prod_{i=1,2} \frac{1}{(k_i^2 - m_W^2)^2}\,
\label{wwcor} 
\eea
and
\bea 
\hspace*{-6mm}
2\, \Re\left[\Delta {\cal M}_W
 \left({\cal M}_Z^{\rm tree}\right)^\dagger\right] 
&=& 4\, C_L^e\, g^{h^0}_{Z Z}\,  \frac{g^4}{\cos^2 \theta_W}
\left(2\,F^{00}\, p_1\cdot p_4\, p_2\cdot p_3 + F^{21}\, S\right)   
\nonumber\\
&\times& \frac{q_2^2 - m_Z^2}{(q_1^2 - m_Z^2)\,
\left((q_2^2 - m_Z^2)^2  + m_Z^2\,\Gamma_Z^2\right)}
\prod_{i=1,2} \frac{1}{k_i^2 - m_W^2}
\, ,
\label{zzcor} 
\eea
where
\bea
S &=& (p_1\cdot p_4 + p_2\cdot p_3)
  \left( p_1\cdot p_2\, p_3\cdot p_4 +
      p_1\cdot p_4\, p_2\cdot p_3  - p_1\cdot p_3\, p_2\cdot p_4\right)
 \nonumber\\
  &-& 2\,(p_1\cdot p_2 + p_3\cdot p_4)\,p_1\cdot p_4\, p_2\cdot p_3 \, ,
\label{sfact}
\eea
$\Gamma_Z$ is the total $Z$-boson width, and
 $C_L^e= -\frac{1}{2} + \sin^2 \theta_W$. The kinematics has been
chosen as 
$e^-(p_1) \; e^+(p_2) \goes \nu_e(p_3) \; \bar{\nu}_e(p_4) \; h^0(p)$. 
We have also defined $k_1=p_3-p_1$, $k_2=p_2-p_4$,
$q_1=p_1+p_2$ and $q_2=p_3+p_4$.

For the calculation of the cross section at 
tree-level, it is possible
to perform some of the phase space integrations analytically and the rest
of them numerically \cite{alta,kilian}. 
However including  the one-loop correction
terms of  Eq.~(\ref{wwcor}) and (\ref{zzcor}), it is
impossible to perform any of these integrations analytically.
Therefore, we have performed  the  integrations
numerically, using efficient numerical integration subroutines
found in the {\small NAG} library.
We have also checked that for the tree-level case our
completely numerical calculation agrees with high
accuracy with the semi-analytical results of Ref.~\cite{kilian}.

Before embarking on  discussing our numerical findings, we will
make some comments concerning some details of our calculation.
For the calculation of the supersymmetric (SUSY)  Higgs boson masses 
and the Higgs mixing angle $\alpha$, a computer  
programme based on  Ref.~\cite{carena} has been used. 
The tree-level $WWh^0$ coupling for values of $\tan\beta > 5$ as preferred
by the LEP Higgs boson searches,  mimics the SM one. 
For the calculation of the fermion/sfermion one-loop
corrections to the $WWh^0$ vertex, the contribution
of the third family of fermions/sfermions has been taken into
account. This contribution turns out to be the dominant one, in
comparison with the first two families corrections, due
to the large values of the Yukawa couplings $h_t$ and $h_b$.
The effect of the running of the coupling constants $g$ and $g'$
has  been taken into account.

\begin{figure}[t]  
\begin{center}
\includegraphics[scale=.75]{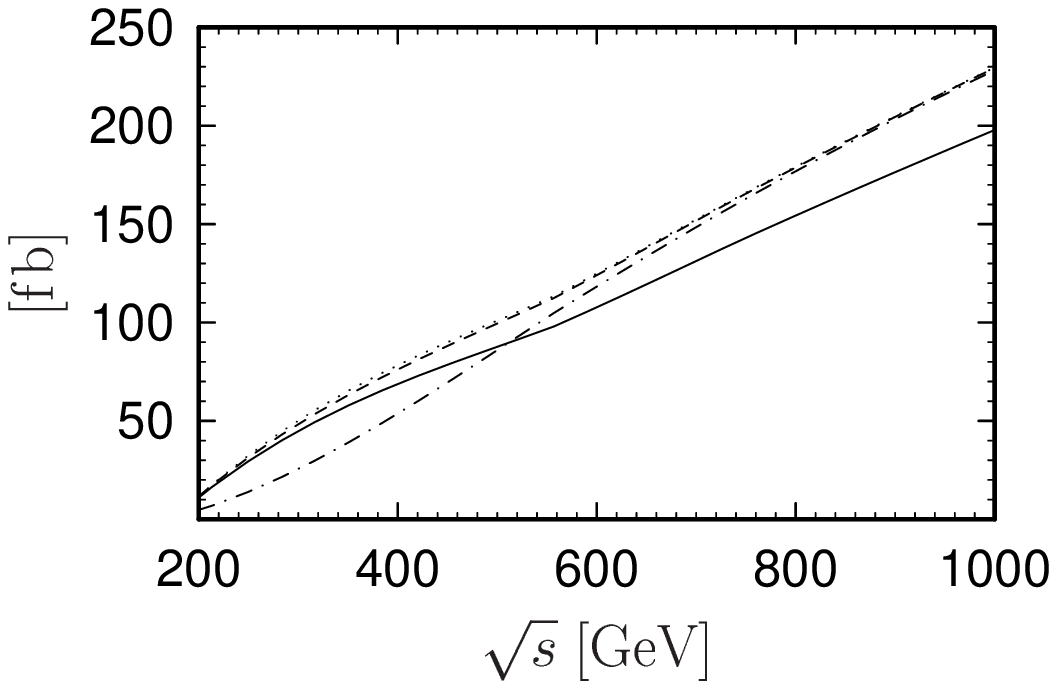}
\includegraphics[scale=.75]{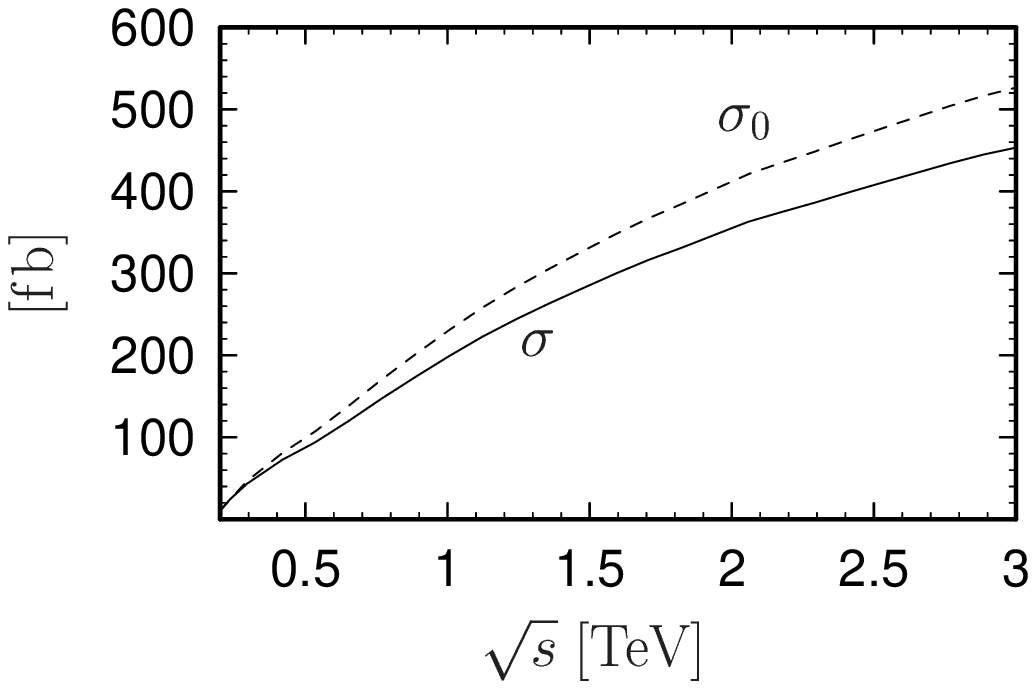} 
\end{center}

\caption[]{The various cross sections as functions of $\sqrt{s}$ (left).
The dotted-dashed line represents the  
tree-level cross section $\sigma_0^{WW}$, 
the dotted line 
$\sigma_0^{WW}+\sigma_0^{h-str}  $. The dashed line 
includes also the interference term $\sigma_0^{\rm interf.}  $ and
represents the total tree-level cross section.
The solid line
includes  the  one-loop correction.
The SUSY parameters are:
$\tan\beta=10$, $\mu=-100\GeV$, $A=-500\GeV$,   
$m_{\tilde{Q}}= 300\GeV$, $M_A=500\GeV$, and $M_2=400\GeV$.
For the same set of parameters also the
tree-level cross section $\sigma_0$ and the one-loop corrected
$\sigma$  are plotted for  $\sqrt{s}$ up to $3\TeV$ (right).
} 
\label{fig3}
\end{figure}

We now turn to the discussion of our numerical results. 
In Fig.~\ref{fig3} (left) we can see the various cross sections
as a function of $\sqrt{s}$ for values up to $1 \TeV$.
The dotted-dashed line represents the contribution from
the $WW$ channel at tree-level alone, whereas the dotted line includes
the Higgsstrahlung contribution as well. The dashed line
comprises in addition the interference between the $WW$ channel and
Higgsstrahlung. One can perceive that the size of this
interference term is extremely small, and for this reason
the difference between the dotted and dashed lines is rather minute. 
It is also clear  that for $\sqrt{s} \gtrsim 500 \GeV$ the
$WW$ fusion contribution  dominates the total cross section
for the Higgs production $\higpro$. Actually, for  
$\sqrt{s} \gtrsim 800 \GeV$  the total tree-level cross section
is due to  $WW$ fusion.
In the  solid line we have taken into account the
one-loop correction from the fermion/sfermion loops.
The correction is always negative
 with a significant  size of the order of $10\%$.
This can also be seen  from  Fig.~\ref{fig3} (right),
where we have plotted the  tree-level cross section $\sigma_0$
(dashed line) and the one-loop corrected
cross section $\sigma$  (solid line) for
energies up to $3 \TeV$. 
For simplicity, for all plots we have used
$A_t=A_b=A_\tau=A$, 
$\{m_{\tilde{U}},m_{\tilde{D}},m_{\tilde{L}},m_{\tilde{E}}\}=$
$\{ \frac{9}{10},\frac{11}{10},1,1 \} \, m_{\tilde{Q}} $
and  $M_1=\frac{5}{3}\, M_2\, \tan^2 \theta_W$.
The choice of a common trilinear coupling and
the correlation between  the soft sfermion masses 
are inspired by unification. 
For  the plots in Fig.~\ref{fig3} we
have taken: 
$\tan\beta=10$, $\mu=-100\GeV$,  $A=-500\GeV$, 
$ m_{\tilde{Q}}= 300\GeV$,
$M_A=500\GeV$, and $M_2=400\GeV$.
Choosing different sets of parameters,
the basic characteristics of these plots remain
indifferent.  
In fact, the soft gaugino masses $M_{1,2}$ affect only the Higgs
sector through radiative corrections.  

\begin{figure}[t]  
\begin{center}
\includegraphics[scale=.75]{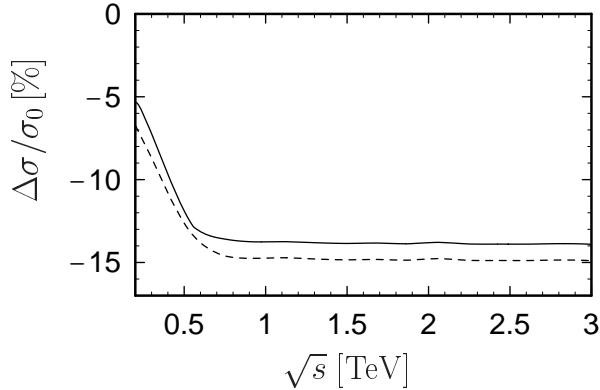} 
\end{center}

\caption[]{The relative correction $\Delta \sigma / \sigma_0$
as a function of $\sqrt{s}$ ($\Delta \sigma= \sigma - \sigma_0$),
where $\sigma_0$ is the tree-level and $\sigma$ the
one-loop corrected cross section.
The solid and dashed lines correspond to two different
choices of the SUSY parameters, as described in the  text.
} 
\label{fig4}
\end{figure}

In  Fig.~\ref{fig4} the relative 
correction $\Delta \sigma / \sigma_0$ is presented as a
function of $\sqrt{s}$ for two different sets of parameters.
The solid line corresponds to the set 
$\tan\beta=10$, $\mu=-100\GeV$, $A=-500\GeV$,  
$ m_{\tilde{Q}}= 300\GeV$, $M_A=500\GeV$, and $M_2=400\GeV$,
 whereas for the
dashed line we have taken  
$\tan\beta=40$, $\mu=-300\GeV$ and $A=-100\GeV$, keeping
the rest of them unchanged.
This figure shows
that the size of the one-loop correction to the
Higgs production cross section is practically
constant for $\sqrt{s} > 500 \GeV$ and weighs about $-15 \%$,
almost independently of the choice  of the
SUSY parameters.
The reason for this weak dependence is that the one-loop
corrections are  dominated by the fermion loops, and
therefore the total correction is not very sensitive
to the choice of the SUSY parameters.
In this region, $\sqrt{s} > 500 \GeV$, one can compare this constant
correction with the effective approximation of Ref.~\cite{kniehl1},
where one only corrects the $WWh^0$ coupling.
Although the sign of this approximation
is correct, it does not fully account for the whole effect.

\begin{figure}[t]  
\begin{center}
\includegraphics[scale=.75]{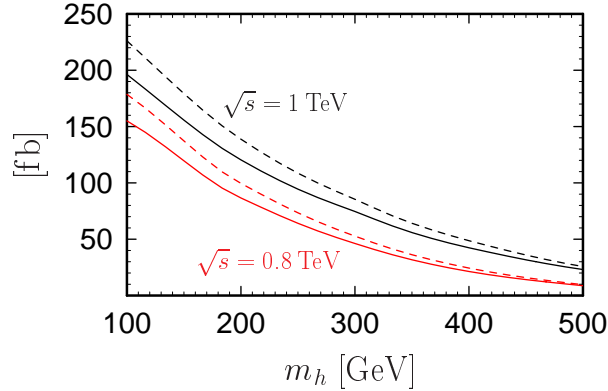} 
\end{center}

\caption[]{Cross sections for the  SM case, 
 for  $\sqrt{s}= 0.8 \TeV$ (red lines) 
and $1\TeV$ (black lines). The dashed lines correspond to
the tree-level cross section, whereas
the solid lines to the one-loop corrected one.
} 
\label{fig5}
\end{figure}

In Fig.~\ref{fig5} we have plotted the cross section 
as a function of $m_h$ for
the SM case, for  $\sqrt{s}= 0.8 \TeV$ (red lines) 
and $1\TeV$ (black lines). The dashed lines correspond to
the tree-level cross section for $\higpro$, whereas
the solid lines contains the one-loop correction stemming
from the fermion loops. In addition, the couplings have
been adjusted to the SM corresponding couplings.   
The plot exhibits the expected dependence of the cross
section on $m_h$. What must be noticed is that
especially for small Higgs boson masses $ \lesssim 200 \GeV$,
the size of the fermion loops correction becomes important
for the correct determination of the Higgs boson mass.
 
\begin{figure}[t]  
\begin{center}
\includegraphics[scale=.75]{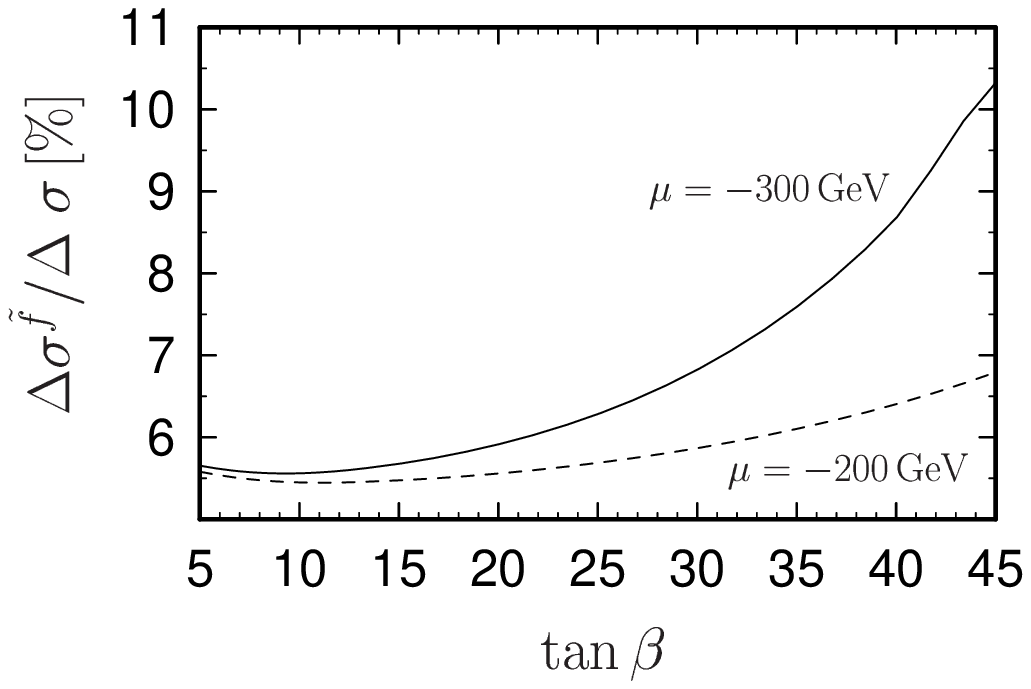}
\includegraphics[scale=.75]{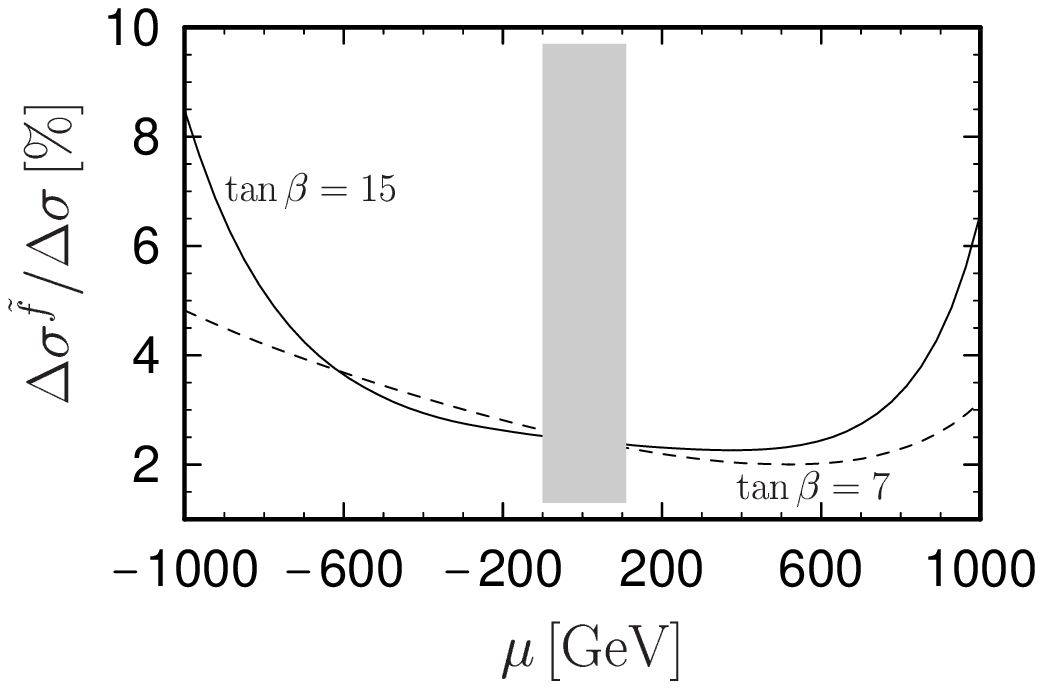} 
\end{center}

\caption[]{The percentage of the sfermions to the total one-loop
correction as a function of $\tan\beta$ (left) and  $\mu$ (right).
The rest of the SUSY parameters have been fixed as described in the
text. Here  $\sqrt{s}=1\TeV$.
The grey area in the right figure is excluded due
to the chargino mass bound.
}
\label{fig6}
\end{figure}

Finally,  Fig.~\ref{fig6} exhibits the percentage of the
sfermion loops to the total one-loop correction as a function
of $\tan\beta$ (left) and $\mu$ (right), for
two different values of $\mu$ and $\tan\beta$, respectively, as
shown in the figure. 
In the left (right) figure we have chosen $A=-100\GeV$ ($A=-400\GeV$).
The rest of SUSY parameters are:
$ m_{\tilde{Q}}= 300\GeV$, $M_A=500\GeV$, and $M_2=400\GeV$.
Here  $\sqrt{s}$ has been fixed to $1\TeV$.
The grey area in the right figure is excluded due
to the chargino mass bound.
 We see that the maximum value
of order 10\% can be achieved for large values of $\mu$ and
$\tan\beta$. There, due to the significant mixing in the stop and sbottom
sector, the  contribution of stops and 
sbottoms in the loops is enhanced. 
For these values of SUSY parameters the sfermion masses approach their
experimental lower bounds.
But even there
the dominant correction, at least  $90\%$ of the 
total correction, is due to the fermion loops.

In conclusion, we have calculated the fermion/sfermion loops
corrections to the single Higgs boson production $\higpro$ in
the context of the MSSM and  SM. 
They are supposed to be the dominant radiative corrections.
For energies relevant to the future
linear colliders, $\sqrt{s} \gtrsim 500 \GeV$, the $WW$ fusion
channel dominates the cross section. 
In general, the  correction due to fermion/sfermion loops is negative 
and yields a correction to the cross section of the order of $-10\%$.
The bulk of this correction stems from the fermion loops, and
usually turns to be more than $90\%$ of the total correction.
For the case of  maximal mixing in the sfermion mass matrices,
the contribution of the sfermion loops is enhanced, but
nevertheless weighs less than $10\%$ of the total one-loop correction.
As the correction is dominated by fermion loops  and is rather independent
of  $\sqrt{s}$ for $\sqrt{s} > 500 \GeV$, 
we think that  it can be approximated by a 
factor correction to the \mbox{tree-level} cross section. Such an
approximation would be most useful for including initial state 
radiation (ISR) and beamstrahlung in an efficient way.

\vspace*{2.5cm}
\noindent 
{\bf Acknowledgements} \\ 
\noindent 
V. C. S. acknowledges support by a Marie Curie Fellowship of the EU
programme IHP under contract HPMFCT-2000-00675.
The authors acknowledge support from
EU under the HPRN-CT-2000-00149 network programme.
The work was also supported by the ``Fonds zur F\"orderung der
wissenschaftlichen Forschung'' of Austria, project No. P13139-PHY.

\end{document}